\documentclass[12pt,aps,prd,superscriptaddress,showpacs,longbibliography,floatfix,nofootinbib]{revtex4-1}

\usepackage[utf8]{inputenc}
\usepackage[OT2,T1]{fontenc}
\usepackage[russian,english]{babel}
\usepackage{slashed}
\pdfoutput=1

\usepackage{color}
\usepackage{graphicx}   
\usepackage{bm}
\usepackage{amsmath}
\usepackage{amsfonts}
\usepackage{hyperref}

\def\be{\begin{equation}}
\def\ee{\end{equation}}
\def\ba{\begin{eqnarray}}
\def\ea{\end{eqnarray}}
\newcommand{\CA}{\mbox{\usefont{OT2}{\rmdefault}{m}{n} D}}
\newcommand{\CQ}{\mbox{\usefont{OT2}{\rmdefault}{m}{n} Q}}
\newcommand{\msbar}{\overline{\rm MS}}

\begin{document}

\title{An alternative to perturbative renormalization in 3+1 dimensional field theories}

\author{Paul Romatschke}
\affiliation{Department of Physics, University of Colorado, Boulder, Colorado 80309, USA}
\affiliation{Center for Theory of Quantum Matter, University of Colorado, Boulder, Colorado 80309, USA}

\begin{abstract}
  Perturbative renormalization provides the bedrock of understanding quantum field theories. In this work, I point out an alternative way of renormalizing quantum field theories, which is naturally encountered and well known for the case of large N scalar field theories. In terms of bare parameters, this non-perturbative alternative renormalization differs qualitatively from its perturbative cousin: in the continuum limit, the bare coupling constant goes to zero instead of infinity, and there is no wave-function counterterm. Despite these differences, the resulting n-point functions of the theory are finite. I provide explicit results for alternative renormalization for the O(N) model and QCD with $N_f=12$ flavors in 3+1 dimensions.
\end{abstract}

\maketitle

\begin{centering}
  Clipped wings, I was a broken thing\\
Had a voice, had a voice but I could not sing\\
You would wind me down\\
I struggled on the ground\\
...\\
But there's a scream inside that we all try to hide\\
We hold on so tight, we cannot deny\\
Eats us alive\\
...\\
And I don't care if I sing off key\\
I find myself in my melodies\\
I sing for love, I sing for me\\
I shout it out like a bird set free\\
...\\
Now I fly, hit the high notes\\
I have a voice, have a voice, hear me roar tonight\\
You held me down\\
But I fought back loud\\
...\\
And I don't care if I sing off key\\
I find myself in my melodies\\
I sing for love, I sing for me\\
I shout it out like a bird set free \cite{Sia}\\
  \end{centering}

\newpage
\section{Introduction}

Renormalization of scalar field theory has been a topic of interest in high energy physics since the dawn of quantum field theory. Regularization and perturbative renormalization for scalar fields is by now standard, and can be found in many variations in various text books \cite{Ramond:1981pw,Ryder:1985wq,Peskin:1995ev}.

For scalar field theory in four dimensions, the perturbative renormalization program starts out the usual way, by first regularizing formally divergent integrals, and then introducing vacuum energy, mass, coupling constant and wave-function counter terms order-by-order in a weak coupling expansion to cancel the divergencies.

In contrast to perturbative approaches, large N expansions of N-component field theories resum an infinite number of perturbative contributions (Feynman diagrams) at every order in $\frac{1}{N}$. This is advantageous in cases where non-perturbative information about the theory is needed, and it has successful applications to different physical systems, see e.g. Ref.~\cite{Romatschke:2023ztk} for a modern review.

However, because large N is not a weak-coupling framework, knowledge of perturbative renormalization of the theory is of little help when aiming for a calculation of renormalized observables at any given order in $\frac{1}{N}$. In practice, one needs to set up non-perturbative renormalization conditions which do not correspond to any finite-order perturbative truncation. At leading order in large N, non-perturbative renormalization is relatively straightforward, and in fact has been used since the 1970s \cite{Abbott:1975bn,Linde:1976qh}. Curiously, Parisi pointed out in a seminal paper that large N field theories may be non-perturbatively renormalized order-by-order in a large N expansion even in cases that are not perturbatively renormalizable \cite{Parisi:1975im}, such as scalar field theory in space-time dimensions $d<6$.

For the case of scalar field theory in $d=4$, the question of non-perturbative renormalization has been encountered in so-called $\Phi$-derivable and 2PI effective theories \cite{vanHees:2001ik,VanHees:2001pf,vanHees:2002bv,Fejos:2009dm}, culminating in the proof of non-perturbative renormalizability in the massless case \cite{Blaizot:2003br,Blaizot:2003an,Berges:2005hc}.

Curiously, there seems to be little discussion about the qualitative differences between non-perturbative and perturbative renormalization in the literature. Specifically for scalar field theory, perturbative renormalization requires adding the counter-terms \cite{Ryder:1985wq}
\be
\delta {\cal L}_V=\chi_V\,,\quad
\delta {\cal L}_\lambda=\chi_\lambda \phi^4, \quad
\delta {\cal L}_m=\chi_m \phi^2, \quad
\delta {\cal L}_\phi=\chi_\phi \partial_\mu \phi \partial_\mu \phi\,,
\ee
to the theory Lagrangian, with the coefficients $\chi_V,\chi_\lambda,\chi_m,\chi_\phi$ all diverging in the continuum limit of the theory. By contrast, as will be reviewed below, non-perturbative renormalization of large N scalar field theory requires adding the counterterms \cite{Romatschke:2022jqg,Romatschke:2023sce}
\be
\delta {\cal L}_V=c_V\,,\quad
\delta {\cal L}_\lambda=c_\lambda \phi^4, \quad
\delta {\cal L}_m=c_m \phi^2\,,
\ee
where in particular $c_\lambda\rightarrow 0$ in the continuum limit of the theory.

One is faced with the situation that while perturbative renormalization requires wave-function renormalization, apparently non-perturbative renormalization does not. In addition, while the coupling constant counterterm in perturbative renormalization diverges in the continuum limit, the corresponding counterterm in non-perturbative renormalization vanishes in the same limit. Yet both renormalization procedures render the theory finite.

I find this difference between perturbative and non-perturbative renormalization striking, and took it as an opportunity to review non-perturbative renormalization in scalar field theory and QCD in this work.


\section{Scalar Field Theory}

The theory is defined by the partition function
\be
Z=\int {\cal D}\vec{\phi} e^{-S_E}\,,
\ee
where $S_E$ is the Euclidean action of the theory
\be
\label{Euclac}
S_E=\int d^4x \left[\frac{1}{2}\partial_\mu \vec{\phi}\partial_\mu \vec{\phi}+\frac{1}{2}m_B^2 \vec{\phi}^2+\frac{\lambda_B}{N}\left(\vec{\phi}^{\ 2}\right)^2\right]\,.
\ee
Here $\vec{\phi}=\left(\phi_1,\phi_2,\ldots,\phi_N\right)$ is a N-component scalar field and $m_B,\lambda_B$ are the bare mass and coupling parameter of the theory. It is useful to rewrite the partition function in terms of an auxiliary field $\zeta$, which is introduced through the mathematical identity
\be
e^{-\int_x \frac{\lambda_B}{N} \left(\vec{\phi}^{\ 2}\right)^2}=\int {\cal D}\zeta e^{-\int_x\left[\frac{i \zeta}{2}\vec{\phi}^2+\frac{\zeta^2 N}{16\lambda_B}\right]}\,.
\ee
It is furthermore convenient to split the auxiliary field into a global zero mode $\zeta_0$ and fluctuations around it
\be
\label{split}
\zeta(x)=\zeta_0+\xi(x)\,,
\ee
such that the partition function becomes
\be
Z=\int d\zeta_0 {\cal D}\vec{\phi} {\cal D}\xi e^{-S_{R0}-S^\prime}\,,
\ee
with
\be
\label{SR0}
S_{R0}=\int d^4x \left[\frac{1}{2}\vec{\phi}\left[-\Box+m_B^2+ i\zeta_0\right] \vec{\phi}+\frac{N}{16\lambda_B}\zeta_0^2\right]\,,\quad
S^\prime=\int d^4x \left[\frac{i \xi \vec{\phi}^2}{2}+\frac{N \xi^2}{16 \lambda_B}\right]\,.
\ee
(Note that the ``Rn'' nomenclature is in one-to-one correspondence to an expansion in powers of $N^{-1}$, cf. Ref.~\cite{Romatschke:2023ztk}).

\subsection{Leading order Large N -- level R0}

So far everything has been exact. Owing to the three vertex $\xi \phi^2$ in $S^\prime$, the path integral for the partition function can not be done in closed form. In the limit of many scalar components $N\gg 1$, the theory is amenable to closed-form solutions in an expansion in $\frac{1}{N}$.

To leading order in large N, the action $S^\prime$ does not contribute to the partition function, and one has
\be
Z_{R0}=\int d\zeta_0 e^{-\frac{N}{2}{\rm Tr}\ln\left[-\Box + m_B^2+i\zeta_0\right]-{\rm vol}\times \frac{N \zeta_0^2}{16\lambda_B}}\,,
\ee
where the Gaussian integral over $\phi$ was performed, and ``${\rm vol}$'' denotes the space-time volume. The functional trace over the logarithm of the operator is UV-divergent, and needs to be regulated. For the present work, I choose to use dimensional regularization, because it is the standard regularization scheme used by the particle data group \cite{Workman:2022ynf}, and because it is elegant and I like elegance. In dimensional regularization, the space-time dimension is taken to be
\be
d=4-2\varepsilon\,,
\ee
with the parameter $\varepsilon>0$ sent to zero in the end. Regularization is provided by the analytic continuation of the $\Gamma$-function, in particular for integrals of the type
\be
\label{masteri}
\int \frac{d^d k}{(2\pi)^d}\frac{1}{(k^2+m^2)^\alpha}=\frac{1}{(4\pi)^{\frac{d}{2}}}
\frac{\Gamma(\alpha-\frac{d}{2})}{\Gamma(\alpha)} (m^2)^{\frac{d}{2}-\alpha}\,.
\ee

For the R0 level partition function, one needs the integral
\ba
\frac{1}{2}{\rm Tr}\ln\left[-\Box + m_B^2+i\zeta_0\right]&=&{\rm vol}\times \frac{1}{2}\int \frac{d^{d}k}{(2\pi)^{d}} \ln\left[k^2+m_B^2+i\zeta_0\right]\,,\nonumber\\
&=&{\rm vol}\times \frac{1}{(4\pi)^{\frac{d}{2}}}
\frac{\Gamma(1-\frac{d}{2})}{d} (m_B^2+i \zeta_0)^{\frac{d}{2}}\,,
\ea
which follows from direct integration of (\ref{masteri}) for $\alpha=1$ with respect to $m^2$.
For $\varepsilon\rightarrow 0$, one therefore finds for the partition function 
\be
\label{ZR0}
Z_{R0}=\int d\zeta_0 e^{-N\times {\rm vol}\times{\cal F}(\zeta_0)}\,,
\ee
with
\be
\label{FR0}
   {\cal F}(\zeta_0)=\frac{\zeta_0^2}{16\lambda_B}-\frac{(m_B^2+i\zeta_0)^2}{64\pi^2}\left[\frac{1}{\varepsilon}+\ln\frac{\bar\mu^2 e^{\frac{3}{2}}}{m_B^2+i\zeta_0}\right]\,,
   \ee
   where $\bar\mu$ is the renormalization scale parameter in the $\overline{\rm MS}$-scheme. 
It is useful to rewrite the expression as
\be
\label{useR0}
      {\cal F}(\zeta_0)=-\frac{(m_B^2+i\zeta_0)^2}{64\pi^2}\left[\frac{4\pi^2}{\lambda_B}+\frac{1}{\varepsilon}+\ln \frac{\bar\mu^2 e^{\frac{3}{2}}}{m_B^2+i\zeta_0}\right]+\frac{m_B^2+i \zeta_0}{32\pi^2}\frac{4\pi^2 m_B^2}{\lambda_B}-\frac{m_B^4}{16\lambda_B}\,.
      \ee

\subsection{Non-perturbative renormalization}
\label{sec:np}

The theory is renormalized if all n-point functions of the theory are finite. Calculating the partition function corresponds to a 0-point function, which means that after integrating over $\zeta_0$, $Z_{R0}$ needs to be finite. In the large N limit, the remaining integral over $\zeta_0$ can be performed exactly using the saddle point method, finding
\be
Z_{R0}=e^{-N\times{\rm vol}\times {\cal F}(\bar \zeta)}\,,
\ee
with ${\cal F}(\zeta_0)$ given by (\ref{useR0}) and $\bar \zeta$ the value of the saddle point. (Note that I'm assuming that there is only one dominant saddle point, which will be verified below). 

      The 2-point function for the R0-level theory is given by
      \ba
      \label{2pointR0}
      \langle \phi_i(x)\phi_j(0)\rangle &=&Z_{R0}^{-1}\int d\zeta_0 {\cal D}\vec{\phi} e^{-S_{R0}} \phi_i(x)\phi_j(0)\,,\nonumber\\
      &=&Z_{R0}^{-1}\int d \zeta_0 e^{-N\times {\rm vol}\times {\cal F}(\zeta_0)}\frac{\delta_{ij}}{-\Box +m_B^2+ i\zeta_0}\,.
      \ea
      This expression involves the two-point function
      \be
      \Delta(x)\equiv \langle x|\frac{1}{-\Box+m_B^2+i \zeta_0}|0\rangle\,,
      \ee
      which is akin to a standard bosonic propagator except for the fact that its ``mass term'' $m_B^2+i \zeta_0$ is still being integrated over. In the large N limit, the integral can again be performed exactly, and one finds
      \be
      \langle \phi_i(x)\phi_j(0)\rangle =\int \frac{d^d k}{(2\pi)^d} \frac{e^{i k\cdot x}}{k^2+m_B^2+i\bar \zeta}\,,
      \ee
      where $\bar \zeta$ is again the saddle point discussed above. As a consequence, the 2-point function of the theory is finite in the large N limit as long as the vector pole mass squared
      \be
      m_{R0}^2=m_B^2+i\bar \zeta\,,
      \ee
      is finite.

      The connected 4-point function for the R0-level theory is given by
      \be
      \langle \phi_a\phi_b\phi_c\phi_d\rangle_{\rm conn.} = Z_{R0}^{-1}\int d\zeta_0 {\cal D}\vec{\phi} {\cal D}\xi e^{-S_{R0}-S^\prime} \phi_a\phi_b\phi_c\phi_d = {\cal O}\left(\frac{1}{N}\right)\,,
      \ee
      to leading order in the large N limit. This is simply a reflection of the fact that non-Gaussian terms do not contribute to the R0-level action to leading order in large N, and hence the 4-point function (as well as all higher order connected n-point functions) are $\frac{1}{N}$ suppressed.

      Therefore, to leading order in large N, the theory is finite if the 0-point and 2-point functions are finite.

      Let's study this explicitly now. In order to renormalize the theory, one can add counterterms to the original action. For the 0-point function, after integration over $\zeta_0$, the requirement that ${\cal F}(\bar \zeta)$ is finite together with the requirement from the 2-point function that $m_{R0}^2$ must be finite implies that the coefficients of the terms $m_{R0}^4,m_{R0}^2,m_{R0}^0$ must be finite. This leads to three renormalization conditions:
      \ba
     \label{renor}
        {\cal O}\left(m_{R0}^4\right)&=&{\rm finite}:\quad {\rm coupling\ renormalization}\,,\\
        {\cal O}\left(m_{R0}^2\right)&=&{\rm finite}:\quad {\rm mass\ renormalization}\,,\\
        {\cal O}\left(m_{R0}^0\right)&=&{\rm finite}:\quad {\rm vacuum\ renormalization}\,,
        \ea
        specifying the form of the coupling-constant, mass and vacuum energy counterterms.

        It is paramount to point out that this does not leave any room for a wave-function renormalization counterterm. Hence I conclude that wave-function renormalization must be absent when non-perturbatively renormalizing large N theories.

        In particular, the counterterm $c_V$ can be used to cancel a potentially divergent vacuum constant in (\ref{useR0}). In the following I choose $c_V$ to fully cancel the term $-\frac{m_B^4}{16\lambda_B}$ in (\ref{useR0}). For the coefficient of $m_{R0}^4$, one finds that it requires adding a counterterm to the \textit{inverse coupling constant} such that
        \be       
        \frac{1}{\lambda_B}=\frac{1}{\lambda_R(\bar\mu)}+c_{\lambda}\,.
        \ee
        Specifically, I take
        \be
        \label{bareR0}
        \frac{1}{\lambda_B}+\frac{1}{4\pi^2 \varepsilon}=\frac{1}{\lambda_R(\bar\mu)}\,,
        \ee
        which renders the ${\cal O}\left(m_{R0}^4\right)$ coefficient in (\ref{useR0}) finite. Finally, the ${\cal O}\left(m_{R0}^2\right)$ coefficient is finite provided that the combination
        \be
        \frac{m_B^2}{\lambda_B}=\frac{m_R^2(\bar\mu)}{\lambda_R(\bar\mu)}\equiv \frac{\alpha_{R0}\Lambda_{\msbar}^2}{4\pi^2}\,,
        \ee
        is finite, which can be accomplished by adding a suitable counterterm $c_m$ to the mass parameter.

        Here $m_R(\bar \mu),\lambda_R(\bar \mu)$ are the (running) renormalized mass and coupling parameter of the theory. From the above conditions, the explicit form of the running parameters is given by
        \be
        \label{RenR0}
      \lambda_R(\bar \mu)=\frac{4\pi^2}{\ln\frac{\Lambda_{\msbar}^2}{\bar \mu^2}}\,,\quad
      m_R^2(\bar \mu)=\frac{\alpha_{R0}\Lambda_{\msbar}^2}{4\pi^2} \lambda_R(\bar \mu)=\frac{\alpha_{R0}\Lambda_{\msbar}^2}{\ln\frac{\Lambda_{\rm R0}^2}{\bar \mu^2}}\,.
      \ee

      The energy scale $\Lambda_{\msbar}$ is an emergent parameter of the theory, corresponding to the $\msbar$ parameter mentioned the Particle Data Book \cite{Workman:2022ynf}. Here it is identical to the so-called Landau pole of the theory. Note that the running coupling approaches zero from below for $\bar\mu \gg \Lambda_{\msbar}$, consistent with the discussions in Ref.~\cite{Romatschke:2022llf,Romatschke:2023sce}.

      As advertised, the counterterms $c_V,c_m,c_\lambda$ are sufficient to render all connected n-point functions of the theory finite in the large N limit. While this finding is trivial to this order in the $\frac{1}{N}$ expansion, I feel compelled to point out the following facts:
      \begin{itemize}
        \item
        The non-perturbative running coupling $\lambda_R(\bar\mu)$ in (\ref{RenR0}) matches the large N part of the 1-loop perturbative running coupling (cf. Ref.~\cite[Eq.(3.1.56)]{makeenko2002methods})
        \be
        \label{lambdapert}
        \lambda_R^{\rm pert}(\bar\mu)=\frac{4\pi^2}{\left(1+\frac{8}{N}\right)\ln\frac{\Lambda_{\msbar}^2}{\bar \mu^2}}\,.
        \ee
      \item
        Sending the regulator $\varepsilon\rightarrow 0$, the bare coupling constant (\ref{bareR0}) in non-perturbative renormalization obeys
        \be
        \label{lambda55}
        \lim_{\varepsilon\rightarrow 0}\lambda_B=\lim_{\varepsilon\rightarrow 0} \frac{1}{\frac{1}{\lambda_R({\bar\mu})}-\frac{1}{4\pi^2 \varepsilon}}\rightarrow 0^-\,,
        \ee
        meaning it approaches zero for any $\lambda_R(\bar\mu)\neq 0$. By contrast, in perturbation theory one \textit{first} expands $\lambda_B$ for $\lambda_R(\bar\mu)\sim 0$ and \textit{then} lets $\varepsilon\rightarrow 0$, leading to
        \be
        \label{pertobs}
        \lim_{\varepsilon\rightarrow 0}\lambda_B^{\rm pert}=\lim_{\varepsilon\rightarrow 0} \lambda_R(\bar\mu)+\frac{\lambda_R^2(\bar \mu)}{4\pi^2 \varepsilon}+{\cal O}\left(\lambda_R^3\right)\rightarrow \infty\,.
        \ee
      \item
        Since non-perturbative renormalization is not linked to a weak-coupling expansion, physical observables are manifestly independent from the (fictitious) renormalization scale $\bar\mu$ order-by-order in $\frac{1}{N}$, as will be shown below.
\end{itemize}

      Let's investigate this last point in more detail by calculating some physical observables from the renormalized large N theory. Inserting the non-perturbative renormalization conditions and explicit forms for $\lambda_R,m_R$ into the expression for ${\cal F}(\zeta_0)$, one finds
      \be
      {\cal F}(\zeta_0)=-\frac{m_{R0}^4}{64\pi^2}\ln \frac{\Lambda_{\msbar}^2e^{\frac{3}{2}}}{m_{R0}^2}+\alpha_{R0} \Lambda_{\msbar}^2\frac{m_{R0}^2}{32\pi^2}\,,
      \ee
      where the I recall that $c_V$ has been chosen to fully cancel the constant in (\ref{useR0}). At large N, the partition function is then determined through its saddle points, and one finds
      \be
      \frac{\ln Z_{R0}}{\rm vol}=-N {\cal F}(\zeta_0=\bar \zeta)=-N \bar {\cal F}\,,
      \ee
      where the saddle-point condition is
      \be
      \label{LOsaddle}
     m_{R0}^2 \ln \frac{\Lambda_{\msbar}^2e^1}{m_{R0}^2}=\alpha_{R0} \Lambda_{\msbar}^2\,.
     \ee
     Depending on the numerical value of $\alpha_{R0}$, there are different solutions to this equation in terms of
     \be
     \label{polemass}
     \bar m_{R0}^2\equiv m_B^2+ i\bar \zeta\,.
\ee
     For $\alpha_{R0}=0$, the solutions are
     \be
     \label{R0saddles1}
     \alpha_{R0}=0: \quad \bar m_{R0}^2=0\,,\quad \bar m_{R0}^2=e^{1}\Lambda_{\msbar}^2\,.
     \ee
     Plugging these values into the partition function, one finds that the trivial solution $\bar m_{R0}^2=0$ corresponds to the trivial vacuum with $\bar {\cal F}=0$, whereas the second solution corresponds to the free energy density solution
     \be
     \label{LOa0}
     \alpha_{R0}=0: \quad \bar {\cal F}=-\frac{e^2\Lambda_{\msbar}^4}{128 \pi^2}\,,
     \ee
     which indicates the thermodynamically preferred phase. This shows that for $\alpha_{R0}=0$, the non-trivial saddle-point solution gives the true large N ground state of the theory.

     For $0<\alpha_{R0}<\alpha_c$, there continue to be two real solution to the saddle-point condition (\ref{LOsaddle}), with the larger value for $\bar m_{R0}^2$ corresponding to the lower-free energy:
     \be
     0<\alpha_{R0}<\alpha_c: \quad \bar m_{R0}^2=-\frac{\alpha_{R0} \Lambda_{\msbar}^2}{W\left(-\frac{\alpha_{R0}}{e}\right)}\,,\quad \bar {\cal F}=-\frac{\bar m_{R0}^4}{128\pi^2}+\frac{\alpha_{R0}  \bar m_{R0}^2\Lambda_{\msbar}^2}{64\pi^2}\,,
     \ee
     with $W(x)$ denoting the Lambert (or product-log) function. The solution stops to be real for $\alpha_{R0}>\alpha_c$, which is located at the branch point of the Lambert function:
     \be
     \alpha_c=1\,.
     \ee
     For the critical value of $\alpha_c$, one has
     \be
     \alpha_{R0}=\alpha_c:\quad \bar m_{R0}^2= \Lambda_{\msbar}^2\,,
     \quad \bar {\cal F}=\frac{\Lambda_{\msbar}^4}{128 \pi^2}\,.
     \ee
     Note the different sign of the free energy for this value of $\alpha_{R0}$ compared to (\ref{LOa0}).

     For decreasing $\alpha_{R0}<0$, the numerical value for the dominant saddle $\bar m_{R0}^2$ (the value of the free energy ${\cal F}$) increases (decreases) monotonically.

     Depending on the numerical value of $\alpha_{R0}$, the vector pole mass changes from
     \be
     \label{LOpolemass}
     \alpha_{R0}=0:\quad \bar m_{R0}=\sqrt{e} \Lambda_{\msbar}\,,\quad
     \alpha_{R0}=\alpha_c:\quad \bar m_{R0}=\Lambda_{\msbar}\,,
     \ee
     e.g. it decreases with increasing $\alpha_{R0}$. Curiously, this implies that irrespective of the mass renormalization, there is a lower limit on the vector mass:
     \be
     \bar m_{R0}\geq \Lambda_{\msbar}\,.
     \ee

     \subsection{Partial Next-to-leading order Large N -- level R1}

     In order to include 1/N corrections in the theory one needs to include additional resummations not covered by the R0 scheme. The simplest such refinement is provided by the so-called R1 resummation scheme outlined and used in Refs.~\cite{Romatschke:2019rjk,Romatschke:2019wxc}. R1 is defined by adding and subtracting a term $\frac{\nu^2}{2}\int d^4x \vec{\phi}^2$ to the action of the theory such that the partition function becomes
     \be
     Z=\int d\zeta_0 {\cal D}\vec{\phi} {\cal D}\xi e^{-S_{R1}-S_{I1}}\,,
     \ee
     with
     \be
     \label{SR1}
S_{R1}=\int d^4x \left[\frac{1}{2}\vec{\phi}\left[-\Box+m_B^2+ i\zeta_0+\nu^2\right] \vec{\phi}+\frac{N}{16\lambda_B}\zeta_0^2\right]\,,\quad
S_{I1}=\int d^4x \left[\frac{i \xi \vec{\phi}^2}{2}+\frac{N \xi^2}{16 \lambda_B}-\frac{\nu^2}{2}\vec{\phi}^2\right]\,,
\ee
where $\nu^2$ is determined by self-consistently calculating the vector two-point function to one-loop level:
\be
\langle \phi_i(x)\phi_j(0)\rangle = \langle \phi_i(x)\phi_j(0)\rangle_{R1}-\langle \phi_i(x)S_{I1} \phi_j(0)\rangle_{R1}+\frac{1}{2}\langle \phi_i(x)S_{I1}^2 \phi_j(0)\rangle_{R1}\,,
\ee
where $\langle \cdot \rangle_{R1}$ specify expectation values for the theory with respect to the action $S_{R1}$. Performing the contractions leads to the result \cite{Romatschke:2019rjk}
\be
\label{nuR1}
\nu^2=\frac{8\lambda_B}{N}\Delta(x=0)\,,
\ee
in terms of the $\zeta_0$-dependent R1 vector propagator
\be
\Delta(x)=\int \frac{d^4k}{(2\pi)^4}\frac{e^{i k_\mu x_\mu}}{k^2+m_B^2+i\zeta_0+\nu^2}\,.
\ee

The R1 partition function is likewise calculated as
\be
Z_{R1}=\int d\zeta_0 e^{-N\times{\rm vol}\times {\cal F}_{R1}(\zeta_0)}\,,
\ee
with the R1-effective potential ${\cal F}_{R1}(\zeta_0)$ given by
\be
{\cal F}_{R1}(\zeta_0)=\frac{\zeta_0^2}{16\lambda_B}+\frac{1}{2}\int \frac{d^4k}{(2\pi)^4}\ln \left[k^2+m_B^2+i\zeta_0+\nu^2\right]-\frac{2\lambda_B}{N} \Delta^2(x=0)\,,
\ee
where for the last term I combined contributions from $S_I$ and $S_I^2$ using (\ref{nuR1}). Calculating the integrals as before using dimensional regularization one finds
\be
{\cal F}_{R1}(\zeta_0)=\frac{\zeta_0^2}{16\lambda_B}-\frac{m_{R1}^4}{64\pi^2}\left(\frac{1}{\varepsilon}+\ln \frac{\bar \mu^2 e^{\frac{3}{2}}}{m_{R1}^2}\right)-\frac{N \nu^4}{32\lambda_B} \,,
\ee
where I introduced the notation
\be
m_{R1}^2=m_B^2+i\zeta_0+\nu^2\,.
\ee

The R1 scheme can be fully non-perturbatively renormalized at any N as follows: 
first note that the R1 partition function in the large volume limit is given by
\be
\frac{\ln Z_{R1}}{{\rm vol}}=-N \bar {\cal F}_{R1}\,,
\ee
with $\bar {\cal F}_{R1}={\cal F}_{R1}(\zeta_0=\bar \zeta)$ and $\zeta_0=\bar \zeta$ the solution to the saddle point condition
\ba
\label{R1saddle}
0=\frac{d{\cal F}}{d (i \zeta_0)}&=&-\frac{i \zeta_0}{8\lambda_B}+\frac{1}{2}\int \frac{d^dk}{(2\pi)^d}\frac{1}{k^2+m_{R1}^2}\left(1+\frac{d\nu^2}{d(i\zeta_0)}\right)-\frac{N \nu^2}{16\lambda_B}\frac{d\nu^2}{d(i\zeta_0)}\,,\nonumber\\
&&=-\frac{i \zeta_0}{8\lambda_B}+\frac{1}{2}\int \frac{d^dk}{(2\pi)^d}\frac{1}{k^2+m_{R1}^2}\,,
\ea
where the last line is a result of (\ref{nuR1}). Comparison of this equation and (\ref{nuR1}) then immediately leads to
\be
\nu^2=\frac{2}{N} i\bar \zeta\,,
\ee
and hence
\be
{\cal F}_{R1}(\bar \zeta)=-\frac{m_{R1}^4}{16\lambda_B \left(1+\frac{2}{N}\right)}+\frac{m_{R1}^2 m_B^2}{8 \lambda_B\left(1+\frac{2}{N}\right)}+\frac{m_B^4}{16\lambda_B \left(1+\frac{2}{N}\right)}-\frac{\bar m_{R1}^4}{64\pi^2}\left(\frac{1}{\varepsilon}+\ln \frac{\bar\mu^2 e^{\frac{3}{2}}}{\bar m_{R1}^2}\right) \,.
\ee
The same steps as for R0 then lead to the renormalization conditions for the coefficients of order $m_{R1}^4,m_{R1}^2,m_{R1}^0$ in the effective potential, finding
\be
\label{renormR1}
\frac{1}{\lambda_B}=\frac{1}{\lambda_R(\bar\mu)}-\frac{1+\frac{2}{N}}{4\pi^2 \varepsilon}\,,\quad \frac{m_B^2}{\lambda_B}=\frac{m_R^2(\bar\mu)}{\lambda_R(\bar\mu)}=\frac{\alpha_{R1}\left(1+\frac{2}{N}\right)\Lambda_{\msbar}^2}{4\pi^2}\,,
\ee
so that in particular the running coupling and mass parameter in the R1 scheme become
\be
        \label{RenR1}
      \lambda_R(\bar \mu)=\frac{4\pi^2}{\left(1+\frac{2}{N}\right)\ln\frac{\Lambda_{\rm \msbar}^2}{\bar \mu^2}}\,,\quad
      m_R^2(\bar \mu)=\frac{\alpha_{R1}\left(1+\frac{2}{N}\right)\Lambda_{\msbar}^2}{4\pi^2} \lambda_R(\bar \mu)=\frac{\alpha_{R1}\Lambda_{\msbar}^2}{\ln\frac{\Lambda_{\msbar}^2}{\bar \mu^2}}\,.
      \ee
Here $\Lambda_{\msbar}$ is an emergent energy scale for the theory that can (once again) be seen to correspond to the $\msbar$ parameter of the theory.
      
Comparing (\ref{RenR1}) to the result from perturbation theory (\ref{lambdapert}) \cite{Ramond:1981pw,Ryder:1985wq,Peskin:1995ev} for $\bar \mu \ll \Lambda_{\msbar}$, one finds that R1 includes \textit{some} of the next-to-leading order corrections, but not all of them. It is also easy to verify that the connected n-point functions with $n\geq 4$ in the R1 level resummation scheme are subleading at large N, so that they are trivially finite. In order to include all NLO large N corrections, the next resummation level needs to be addressed.

Before doing so, let's again calculate some observables. Choosing again $c_V$ to cancel the constant term $m_{R1}^0$ in ${\cal F}$, one has
\be
\bar {\cal F}_{R1}=-\frac{\bar m_{R1}^4}{64\pi^2}\ln \frac{\Lambda_{\msbar}^2 e^{\frac{3}{2}}}{\bar m_{R1}^2}+\alpha_{R1}  \Lambda_{\msbar}^2\frac{\bar m_{R1}^2}{32\pi^2}\,,
\ee
where $\bar m_{R1}^2$ is given by the saddle point condition (\ref{R1saddle})
\be
\label{pNLOsaddle}
m_{R1}^2\ln \frac{e^1 \Lambda_{\msbar}^2}{m_{R1}^2}=\alpha_{R1}\Lambda_{\msbar}^2\,,
\ee
cf. Eq.~(\ref{LOsaddle}). In terms of the parameters $\Lambda_{\msbar},\alpha_{R1}$, the vector pole mass and the free energy take on the same values as for the leading-order large N expansion, e.g.
\be
\alpha_{R1}=0:\quad \bar m_{R1}^2=e^1 \Lambda_{\msbar}^2\,, \quad \bar {\cal F}_{R1}=-\frac{e^2 \Lambda_{\msbar}^4}{128\pi^2}\,.
\ee

     \subsection{Next-to-leading order Large N -- level R2}

     Complete corrections to subleading order in large N are effectively captured by employing the R2 level resummation of (\ref{SR0}), cf. \cite{Romatschke:2019rjk,Romatschke:2021imm,Weiner:2022kgx} for details. To this end, add and subtract self-energy terms for the vector and auxiliary field in the action, obtaining
     \be
     S_{R0}+S^\prime={\rm vol}\times \frac{N \zeta_0^2}{16\lambda_B}+S_{R2}+S_{I2}\,,
     \ee
     where
     \ba
     S_{R2}&=&\int d^dx d^dy\frac{1}{2}\left[\vec{\phi}(x)\Delta^{-1}(x-y)\vec{\phi}(y)+\xi(x)D^{-1}(x-y)\xi(y)\right]\,,\\
     S_{I2}&=&\int d^dx \frac{i \xi \vec{\phi}^2}{2}-\frac{1}{2}\int d^dx d^dy\left[ \vec{\phi}(x)\Sigma(x-y)\vec{\phi}(y)+N \xi(x)\Pi(x-y) \xi(y)\right]\,,
     \ea
     and in Fourier space
     \be
     \Delta(k)=\frac{1}{k^2+m_B^2+i\zeta_0+\Sigma(k)}\,,\quad D(k)=\frac{1}{N}\frac{1-\frac{\delta(k)}{\rm vol}}{\frac{1}{8\lambda_B}+\Pi(k)}\,,
     \ee
     where the $\delta$-function in $D(k)$ is a consequence of the constraint $\int d^d x \xi(x)=0$, cf. (\ref{split}) and the discussion in Ref.~\cite{Weiner:2022kgx}.

     The self-energies $\Sigma,\Pi$ are found by calculating the correlators $\langle \phi(x)\phi(0)\rangle$, $\langle \xi(x) \xi(0)\rangle$ self-consistently to one-loop level, finding \cite{Romatschke:2019rjk,Romatschke:2023ztk}
     \be
     \Pi(x)=\frac{1}{2}\Delta^2(x)\,,\quad \Sigma(x)=D(x) \Delta(x)\,.
     \ee
     Because $D(x)\propto \frac{1}{N}$, at large N the self-energy obeys $\Sigma(x)\propto \frac{1}{N}$, and I do not need to include the self-energy $\Sigma$ when evaluating $\Pi(x)$ to leading order in large N:
     \be
     \Pi(k)=\frac{1}{2}\int \frac{d^d p}{(2\pi)^d} \frac{1}{\left((k-p)^2+m_B^2+i\zeta_0\right)\left(p^2+m_B^2+i\zeta_0\right)}\,.
     \ee
     Using Feynman parameters and shifting the momentum variable, this becomes
     \be
     \Pi(k)=\frac{1}{2}\int_0^1 dx\int \frac{d^d p}{(2\pi)^d} \frac{1}{\left(p^2+m_B^2+i \zeta_0+k^2 x(1-x)\right)^2}\,,
     \ee
     which contains a UV divergence in the momentum integral. Regulating the integral once again using dimensional regularization, one finds
     \be
     \Pi(k)=\frac{1}{32\pi^2}\left[\frac{1}{\varepsilon}+\ln \frac{\bar \mu^2 e^{2}}{m_B^2+i\zeta_0}-2 \sqrt{\frac{k^2+4 (m_B^2+i \zeta_0)}{k^2}}{\rm atanh}\sqrt{\frac{k^2}{k^2+4 (m_B^2+i \zeta_0)}}\right]\,.
     \ee
     The auxiliary field propagator $D(k)$ is already ${\cal O}\left(\frac{1}{N}\right)$, so $\frac{1}{N}$ corrections to the combination $m_B^2+i\zeta_0=m_{R0}^2$ and bare coupling constant $\lambda_B$ can be neglected. Hence using (\ref{bareR0}), the auxiliary field propagator then becomes
     \be
     \label{R2prop}
     D(k)=\frac{32\pi^2}{N}\frac{1-\frac{\delta(k)}{\rm vol}}{\ln \frac{\Lambda_{\msbar}^2e^{2}}{m_{R0}^2}-2 \sqrt{\frac{k^2+4 m_{R0}^2}{k^2}}{\rm atanh}\sqrt{\frac{k^2}{k^2+4 m_{R0}^2}}}\,.
     \ee

     Explicitly, using (\ref{LOpolemass}) one has
     \be
     \alpha_{R0}=0:\quad  D(k)=\frac{32\pi^2}{N}\frac{1-\frac{\delta(k)}{\rm vol}}{1-2 \sqrt{\frac{k^2+4 \bar m_{R0}^2}{k^2}}{\rm atanh}\sqrt{\frac{k^2}{k^2+4 \bar m_{R0}^2}}}\,,
     \ee
     and
     \be
     \alpha_{R0}=\alpha_c:\quad  D(k)=\frac{32\pi^2}{N}\frac{1-\frac{\delta(k)}{\rm vol}}{2-2 \sqrt{\frac{k^2+4 \bar m_{R0}^2}{k^2}}{\rm atanh}\sqrt{\frac{k^2}{k^2+4 \bar m_{R0}^2}}}\,,
     \ee
     which are well defined  and finite for any real-valued Euclidean momentum $k_\mu$. However, note that for $\alpha=\alpha_c$, the zero in the denominator for $k=0$ only does not lead to a pole in $D(k)$ because of the vanishing numerator for this case.

     In the following, for simplicity I will limit myself to only considering the case
     \be
     \label{choice}
     \alpha_{R0}=0\,.
     \ee
    
     The vector self-energy complete to next-to-leading order in large N is then given as
     \be
     \Sigma(k)=\int \frac{d^d p}{(2\pi)^d} \frac{D(p)}{(k-p)^2+m_{R0}^2}\,.
     \ee
     Formally expanding out $\Sigma(k)$ in perturbation theory, one finds that it contains terms such as the ``setting-sun'' diagram, which are known to contain UV divergencies proportional to $k^2$, e.g. those that require wave-function renormalization in perturbation theory \cite{Arjun:2021gaf}.

     For the case at hand, it is easy to check if $\Sigma(k)$ contains UV divergencies proportional to $k^2$. To this end, consider
     \be
     \label{finite}
     \frac{\partial}{\partial k_\mu} \frac{\partial}{\partial k_\mu} \Sigma(k)=- 8 m_{R0}^2
     \int \frac{d^d p}{(2\pi)^d} \frac{D(p) }{\left[(k-p)^2+m_{R0}^2\right]^3}\,.
     \ee
     Since $D(p)<p^2$ for large momenta, the resulting expression is UV finite. Hence, there are no momentum-dependent divergencies in $\Sigma(k)$, unlike what happens in almost all orders of a perturbative expansion of $\Sigma(k)$.

     In fact, (\ref{finite}) is UV-finite at every value of $k$, which implies that any potential UV divergence of $\Sigma(k)$ must be contained in
     \be
     \label{s0}
     \Sigma(0)=\int \frac{d^d p}{(2\pi)^d} \frac{D(p)}{p^2+m_{R0}^2}\,,
     \ee
     and that $\Delta \Sigma(k)\equiv \Sigma(k)-\Sigma(0)$ must be UV finite to next-to-leading order in the large N expansion.  Note that this differs radically from usual perturbative renormalization. 

     In the R2-level resummation, the partition function is given by
     \be
     Z_{R2}=\int d\zeta_0 e^{-N\times{\rm vol}\times {\cal F}_{R2}(\zeta_0)}\,,
     \ee
     with the R2-level free energy density ${\cal F}_{R2}(\zeta_0)$ given by \cite{Romatschke:2019rjk}
     \ba
     \label{FR2level}
        {\cal F}_{\rm R2}(\zeta_0)&=&\frac{\zeta_0^2}{16\lambda_B}+\frac{1}{2}\int \frac{d^dk}{(2\pi)^d} \ln \left[k^2+m_{R0}^2+\Sigma(k)\right]-\frac{1}{2}\int \frac{d^dk}{(2\pi)^d} \Sigma(k)\Delta(k)\nonumber\\
        &&+\frac{1}{2 N}\int \frac{d^d k}{(2\pi)^d}\ln\left[D^{-1}(k)\right]\,,
        \ea
        where a two-loop diagram has canceled. Noting that $\Sigma(k)\propto \frac{1}{N}$, one can expand the first logarithm in ${\cal F}_{\rm R2}$ for large N, finding that the total contribution of $\Sigma$ is ${\cal O}(N^{-2})$ , or next-to-next-to-leading order in large N, which is beyond the accuracy of this work.
        Hence at next-to-leading order in large N, the expression for the free energy functional simplifies to
        \be
        \lim_{N\gg 1} {\cal F}_{\rm R2}[\zeta_0]=\frac{\zeta_0^2}{16\lambda_B}+\frac{1}{2}\int \frac{d^dk}{(2\pi)^d} \ln \left[k^2+m_{R0}^2\right]+\frac{1}{2 N}\int \frac{d^dk}{(2\pi)^d}\ln\left[D^{-1}(k)\right]\,.
        \ee
        All but the last integral in this expression have been evaluated in the R0 subsection above.

        To extract the divergencies of the functional trace over the auxiliary field propagator, note that in dimensional regularization only logarithmic\footnote{An earlier version of this work treated this dogmatically, extracting only divergencies that register as $\frac{1}{\varepsilon}$ in dimensional regularization. A more careful analysis shows that besides divergencies of this form, also terms such as $\ln \varepsilon$ can appear, corresponding to $\ln \ln \Lambda_{\rm UV}$ divergencies in cut-off regularization.} divergencies register. As a consequence, expanding $\ln D^{-1}(k)$ for $k\gg m_{R0}$,
        the divergent piece of the integral must originate from the term proportional to $k^{-4}$ in this power series expansion. One finds
        \be
        \ln D^{-1}(k)=\ldots - \frac{4 m_{R0}^4}{k^4}+\frac{\left(3+6\ln \frac{\Lambda_{\overline{\rm MS}}^2 e^2}{m_{R0}^2}\right) m_{R0}^4}{k^4\ln \frac{m_{R0}^2}{k^2}}+ \ldots\,,
        \ee
        and hence
        \be
        \frac{1}{2N}\int \frac{d^dk}{(2\pi)^d}\ln\left[D^{-1}(k)\right]=-\frac{ m_{R0}^4}{N}\frac{1}{8\pi^2\varepsilon}+\frac{m_{R0}^4}{N}\frac{3+6\ln \frac{\Lambda_{\overline{\rm MS}}^2 e^2}{m_{R0}^2}}{32\pi^2}\ln\varepsilon+{\rm finite}\,.
        \ee
        Collecting the leading order large N result and the divergent pieces and bare parameter contributions in ${\cal F}_{R2}$ one finds
        \be
        \label{fnlofull}
        \lim_{N\gg 1} {\cal F}_{\rm R2}(\zeta_0)=-\frac{(i\zeta_0)^2}{16\lambda_B}
        -\frac{m_{R0}^4}{64\pi^2}\left[\frac{1}{\varepsilon}\left(1+\frac{8}{N}\right)+\ln \frac{\bar \mu^2 e^{\frac{3}{2}}}{m_{R0}^2}-\frac{12\ln \frac{\Lambda_{\overline{\rm MS}}^2 e^\frac{5}{2}}{m_{R0}^2}}{N}\ln \varepsilon\right]+{\rm finite}\,.
        \ee

        Because of the choice (\ref{choice}), the bare mass parameter vanishes to leading order in large N, so one can set $i \zeta_0=m_{R0}^2$. According to the principles of non-perturbative renormalization set out in section \ref{sec:np}, the coefficients of the terms $m_{R0}^4,m_{R0}^2,m_{R0}^0$ have to be finite. However, inspecting (\ref{fnlofull}), the term proportional to $ m_{R0}^4 \ln m_{R0}^2\ln \varepsilon$ does not fit into this scheme: since the logarithm depends on $m_{R0}$, no single choice of $\lambda_B$ can be used to absorb this divergence for all $m_{R0}$.

        However, there is a way to render ${\cal F}_{R2}$ finite. In fact, it will become clear that this is a required operation in order to obtain a finite two-point function for the theory. Namely, rescaling $i\zeta_0$ by a (divergent) constant as
        \be
        i\zeta_0\rightarrow i\zeta_0\left(1+\frac{c_\Sigma}{N}\right)\,,
        \ee
        and expanding ${\cal F}$ to NLO in large N leads to
        \ba
        \label{fnlo2}
        \lim_{N\gg 1} {\cal F}_{\rm R2}(\zeta_0\left(1+\frac{c_\Sigma}{N}\right))&=&
        -\frac{(i\zeta_0)^2}{64\pi^2}\left[\frac{4\pi^2}{\lambda_B}+\ln \frac{\bar \mu^2 e^{\frac{3}{2}}}{i \zeta_0}+\frac{1}{\varepsilon}\left(1+\frac{8}{N}\right)
         \right.\nonumber\\
          &&\left.
           +\frac{2 c_\Sigma}{N}\left(\frac{4\pi^2}{\lambda_B}+\ln \frac{\bar \mu^2 e^{1}}{i \zeta_0}+\frac{1}{\varepsilon}\right)-\frac{12\ln \frac{\Lambda_{\overline{\rm MS}}^2 e^\frac{5}{2}}{i \zeta_0}}{N}\ln \varepsilon\right]+{\rm finite}\,.
        \ea
        The term proportional to $c_\Sigma$ is already ${\cal O}\left(N^{-1}\right)$, so that the leading order renormalization (\ref{bareR0}) can be used without making any error to NLO in large N. One finds
\ba
        \label{fnlo3}
        \lim_{N\gg 1} {\cal F}_{\rm R2}(\zeta_0\left(1+\frac{c_\Sigma}{N}\right))&=&
        -\frac{(i\zeta_0)^2}{64\pi^2}\left[\frac{4\pi^2}{\lambda_B}+\ln \frac{\bar \mu^2 e^{\frac{3}{2}}}{i \zeta_0}+\frac{1}{\varepsilon}\left(1+\frac{8}{N}\right)
         \right.\nonumber\\
          &&\left.
           +\frac{2 c_\Sigma}{N}\ln \frac{\bar \Lambda_{\overline{\rm MS}}^2 e^{1}}{i \zeta_0}-\frac{12\ln \frac{\Lambda_{\overline{\rm MS}}^2 e^\frac{5}{2}}{i \zeta_0}}{N}\ln \varepsilon\right]+{\rm finite}\,.
        \ea
        It is now apparent that the offending term can be canceled by choosing
        \be
        \label{csigma}
        c_\Sigma=6 \ln \varepsilon\,,        
        \ee
        so that with this choice
        \be
         \label{fnlo4}
        \lim_{N\gg 1} {\cal F}_{\rm R2}(\zeta_0\left(1+\frac{c_\Sigma}{N}\right))=
        -\frac{(i\zeta_0)^2}{64\pi^2}\left[\frac{4\pi^2}{\lambda_B}+\ln \frac{\bar \mu^2 e^{\frac{3}{2}}}{i \zeta_0}+\frac{1}{\varepsilon}\left(1+\frac{8}{N}\right)
         -\frac{18}{N}\ln \varepsilon\right]+{\rm finite}\,.
        \ee

        In this form now, the free energy can be rendered finite by a non-perturbative coupling constant renormalization:        
        \be
        \label{renormR2}
        \frac{1}{\lambda_B}=\frac{1}{\lambda_R(\bar\mu)}-\frac{1+\frac{8}{N}}{4\pi^2 \varepsilon}+\frac{9\ln \varepsilon}{2\pi^2 N}\,.
        \ee

        Replacing $\ln \varepsilon\rightarrow \ln \left(-\frac{\lambda_B}{4\pi^2}\right)+{\cal O}(N^{-1})$, the above renormalization is qualitatively similar to the form expected from two-loop running in perturbation theory.

        The non-perturbative renormalization condition (\ref{renormR2}) renders finite ${\cal F}_{\rm R2}$ in the large N limit, which in turn gives a finite value for the saddle $\bar \zeta$ which in turn guarantees a finite 0-point function to next-to-leading order in large N (after subtracting a divergent vacuum energy contribution). One can check this explicitly by re-calculating the saddle-point condition directly from (\ref{FR2level}):
        \be
        \label{saddleR2}
        0=\frac{d {\cal F}_{\rm R2}}{d i\zeta_0}=-\frac{i \bar\zeta}{8\lambda_B}\left(1+\frac{c_\Sigma}{N}\right)+\frac{1}{2}\int \frac{d^dk}{(2\pi)^d} \frac{1}{k^2+i \bar\zeta\left(1+\frac{c_\Sigma}{N}\right)+\Sigma(k)}\,,
        \ee
        where several terms have canceled, cf. Ref.~\cite{Romatschke:2019rjk}. At this point it is useful to recall that the 2-point function for the vector fields of the theory is given by
        \be
        \label{NLO2point}
      \langle \phi_i(x)\phi_j(x)\rangle=\int \frac{d^dk}{(2\pi)^d}
      \frac{e^{i k x}}{k^2+i \bar \zeta\left(1+\frac{c_\Sigma}{N}\right)+\Sigma(k)}\,,
      \ee
      where $\bar \zeta$ is finite after renormalizing the coupling (\ref{renormR2}). I have pointed out above that $\Sigma(k)$ does not contain any divergencies proportional to $k^2$. It would seem that a divergent constant $\Sigma(0)$ would lead to a divergence in the two-point function. The only way to avoid a divergent pole mass is therefore that the divergence in $\Sigma(0)$ is exactly canceled by the choice of $c_\Sigma$ that was required to render the free energy finite for all $i\zeta_0$.

      It should be stressed that the term $c_\Sigma$, though divergent, does not correspond to a wave-function counterterm. This is because a wave-function counterterm is very specific: it is an additional term $\chi_\phi \partial_\mu \vec{\phi}\cdot \partial_\mu \vec{\phi}$ in the Euclidean action (\ref{Euclac}), which manifests itself as an additional term $2 k^2 \chi_\phi$ in the denominator of (\ref{NLO2point}). By contrast, the term proportional to $c_\Sigma$ in (\ref{NLO2point}) is independent of the momentum $k$, hence $c_\Sigma$ cannot be a wave-function counterterm. In fact, from its origin, it is clear that $c_\Sigma$ is no counterterm at all, because it does not correspond to an additional term in the action (\ref{Euclac}). There is no additional renormalization condition associated with $c_\Sigma$: it simply has to cancel out against the corresponding divergent part of $\Sigma(0)$.
      
      After this discussion, let's calculate the divergent part of $\Sigma(0)$: expanding $\frac{D(p)}{p^2+m_{R0}^2}$ in (\ref{s0}) in powers of momenta $p \gg m_{R0}$, any logarithmic divergence must originate from the term proportional to $p^{-4}$ in this power series expansion. One finds that
      \be
        \frac{D(p)}{p^2+m_{R0}^2}=\frac{32\pi^2}{N}\left[\ldots - \frac{3 m_{R0}^2}{p^4 \ln \frac{m_{R0}^2}{p^2}}+ \ldots\right]\,,
        \ee
        so that
        \be
        \Sigma(0)=-\frac{6 m_{R0}^2}{N}\ln \varepsilon+{\rm finite}\,,
        \ee
        confirming that $\Sigma(0)$ is indeed divergent in the limit $\varepsilon\rightarrow 0$.

        From (\ref{NLO2point}), the inverse propagator becomes
        \be
        k^2+i\bar \zeta\left(1+\frac{c_\Sigma}{N}\right)+\Sigma(0)+\Delta\Sigma(k)=k^2+i\bar \zeta+\Delta\Sigma(k)+{\rm finite}\,,
        \ee
        because to this order in the large N expansion and using (\ref{csigma})
        \be
        i\bar \zeta \frac{c_\Sigma}{N}=\frac{m_{R0}^2 c_\Sigma}{N}=\frac{6 m_{R0}^2 \ln \varepsilon}{N}
        \ee
where I recall that $\Delta \Sigma(k)=\Sigma(k)-\Sigma(0)$ was found to be UV-finite above.

The only divergence present in $\Sigma(k)$ is precisely canceled by the shift choice (\ref{csigma}) which was necessary to cancel the logarithmic term in the free energy.

        As a consequence, the explicit construction above shows that both the 0-point and 2-point function of the theory are finite to NLO in the large N expansion.
        
        The connected 4-point function for the R2-level theory is given by
      \be
      \langle \phi_a\phi_b\phi_c\phi_d\rangle_{\rm conn.} = Z_{R2}^{-1}\int d\zeta_0 {\cal D}\vec{\phi} {\cal D}\xi e^{-S_{R2}-S_{I2}} \phi_a\phi_b\phi_c\phi_d\,,
      \ee
      which allows for different kinematic channels (``s'', ``t'' and ``u'' channel), all of which have the same form. For instance, in the s-channel, one finds in momentum space for the R2-level connected amputated 4-point function
      \be
      \Gamma_4^{(s)}(p)=\frac{1}{32\pi^2}D(p)=\frac{1-\frac{\delta(p)}{\rm vol}}{\frac{N}{8\lambda_B}+\Pi(p)}\,.
      \ee
      One can immediately see that $\Gamma_4(p)\propto \frac{1}{N}$, which is why one needs to go to at least the R2 resummation level in order to get a consistent result for the 4-point function.

      One can also see that $\Gamma_4(p)$ is amenable to a perturbative expansion
      \be
      \Gamma_4^{(s)}(p\neq 0)=\frac{8\lambda_B}{N}-\frac{(8 \lambda_B)^2 \Pi(p)}{N^2}+\ldots\,,
      \ee
      where in perturbation theory every single term is divergent because of (\ref{pertobs}), cf. the discussion in Ref.~\cite{Romatschke:2023ztk}. In strong contradistinction, at the R2 level in the large N approximation $\Gamma_4^{(s)}(p)$ is given exactly as
\be
      \Gamma_4^{(s)}(p)=\frac{1}{N}\frac{1-\frac{\delta(p)}{\rm vol}}{1-2 \sqrt{\frac{p^2+4 \bar m_{R0}^2}{p^2}}{\rm atanh}\sqrt{\frac{p^2}{p^2+4 \bar m_{R0}^2}}}\,,
      \ee
      where I used again (\ref{R2prop}) for $\alpha_{R0}=0$. This expression for the 4-point function of the theory is finite and well-behaved for all Euclidean momenta $p$.

It is possible to consider the connected 6-point function for the theory, but one finds that it is proportional to $\frac{1}{N^2}$ so that the R2-level resummation scheme is insufficient to calculate this quantity. While it is possible to study 6-point functions using higher order resummation schemes, such as in particular R4 \cite{Romatschke:2021imm}, this calculation is left for future work.

\subsection{Summary of non-perturbative renormalization for scalar field theory}

To summarize, it is possible to set up an alternative renormalization scheme in scalar field theories different from perturbative renormalization that leads to finite n-point functions order-by-order in a large N expansion.

This alternative renormalization scheme does not allow for wave-function renormalization, which is different from the perturbative approach.

Also, in the non-perturbative renormalization approach, the bare coupling parameter tends to $\lambda_B\rightarrow 0^-$ as the regularization parameter is removed, which is qualitatively different from the perturbative behavior.

Despite these unusual features, and in particular despite employing fewer counterterms than perturbative renormalization, I provided explicit calculations valid up to and including ${\cal O}(N^{-1})$ to show that all connected n-point functions of the theory come out finite. In particular, I showed that the connected 4-point function to order ${\cal O}\left(\frac{1}{N}\right)$ is finite even though its perturbative re-expansion is divergent at every single order in perturbation theory.

I find it intriguing that the theory, defined through a controlled expansion in $\frac{1}{N}$, is apparently ``less divergent'' than its perturbative definition, by which I mean that in the massless case a single coupling-constant renormalization is sufficient in dimensional regularization to render all n-point functions finite to order ${\cal O}\left(\frac{1}{N}\right)$, whereas perturbative renormalization requires both coupling constant and wave-function counterterms.

In my opinion, this provides further evidence for the claim that a non-trivial interacting scalar field theory formulation is possible in the continuum, cf. Ref.~\cite{Romatschke:2023sce}. Also, applications of this alternative renormalization, for instance to recent strong coupling real-time problems such as in Refs.~\cite{Heller:2023mah,Gelis:2023bxw} are possible.

Another possible application of the above findings is to try to make contact with lattice scalar field theory, which constitutes its own non-perturbative renormalization scheme. A direct comparison requires lattice scalar field theory in four dimensions with bare couplings that fulfill (\ref{lambda55}), which is a study in its infancy \cite{Lawrence:2022afv,Romatschke:2023fax}. However, in \textit{two dimensions}, where lattice results only require non-perturbative mass renormalization, the non-perturbative lattice renormalization and the non-perturbative renormalization discussed in this work are \textit{identical}, cf. Refs.~\cite{Schaich:2009jk,Romatschke:2019rjk}.

\section{QCD}

Inspired by the success of non-perturbative renormalization in scalar field theory, let me now consider a similar set-up for QCD. Unlike in the case of scalar field theory, there is no simplification in performing a large $N_c$ expansion for QCD. Instead, I will be using a 'semi-classical' expansion of the QCD effective action, which can be formalized as a power expansion in $\hbar$. However, since $\hbar$ is not dimensionless, such an expansion is necessarily uncontrolled, which means that the results presented in the following are qualitative in nature. Nevertheless, I hope that some readers will find the apparent similarity with the scalar field theory case amusing.

Specifically, following the work reported in Ref.~\cite{Grable:2023eyj}, I consider an SU(N) gauge theory coupled to $N_f$ flavors of massless Dirac spinors in the fundamental representation with an action given by
\be
\label{SQCD}
S_E=\int d^dx \left[\frac{1}{4 g_B^2} G_{\mu\nu}^a G_{\mu\nu}^a+\bar\psi_f \slashed{D} \psi_f\right]
\ee
where $g_B$ is the bare Yang-Mills coupling constant. Here
\be
\label{fmunu}
G_{\mu\nu}^a=\partial_\mu A_\nu^a-\partial_\nu A_\mu^a+f^{abc}A_\mu^b A_\nu^c
\ee
is the Yang-Mills field strength tensor, $A_\mu^a$ is the gauge potential, $\psi_f=\left(\psi_1,\psi_2,\ldots,\psi_{N_f}\right)$ are the $N_f$ Dirac fermions and $\slashed{D}=\gamma_\mu \left(\partial_\mu-i A_\mu^a t^a\right)$ is the Dirac operator with the Euclidean gamma matrices $\gamma_\mu$ with commutation relation $\left\{\gamma_\mu,\gamma_\nu\right\}=2 \delta_{\mu\nu}$. The SU(N) generators $t^a$ and structure constants $f^{abc}$ are taken to fulfill
\be
   {\rm tr}\, t^{a}t^{b}=\frac{\delta^{ab}}{2}\,,\quad \left[t^a,t^b\right]=i f^{abc}t^c\,.
   \ee

   Inspecting the QCD action (\ref{SQCD}) and comparing it to the scalar field action in the auxiliary field formulation (\ref{SR0}), one can observe certain similarities. For instance, the ``matter content'' given by the scalars $\phi_i$ and the fermions $\psi_f$, respectively, appears quadratically in both cases. Similarly, the matter fields couple through a three-point vertex to the auxiliary field $\zeta$ for scalars, and through a three-point vertex to the gauge field $\gamma_\mu A_\mu^a t^a$ in QCD. For the case of scalar field theory, the leading large N partition function could be isolated by splitting the auxiliary field into a constant plus fluctuations, e.g. (\ref{split}). This motivates performing a similar split of the gauge field in QCD as 
\be
\label{splitQCD}
A_\mu^a(x)=\CA_\mu^a(x)+a_\mu^a(x)\,,
\ee
where $\CA_\mu^a(x)$ is referred to the ``background field'' and can be an arbitrary (but fixed) function of coordinates. The QCD partition function then becomes
\be
Z=\int d \CA {\cal D} a {\cal D}\bar\psi {\cal D}\psi e^{-S_E}\,,
\ee
where $\int d\CA$ is a single integral (not a path integral) over the fixed background function $\CA_\mu^a$. Expanding the QCD action in powers of the fluctuation field $a_\mu^a$
as
\be
S_E=S_0[\CA]+S_1[\CA]+S_2[\CA]+\ldots\,,
\ee
with $S_n[\CA]$ corresponding to the contribution of n powers of $a_\mu^a$ in the expansion of the action. In particular, one has
\be
S_2[\CA]=\frac{1}{4 g_B^2}\int d^d x\left[\left(d_\mu^{ac}a_\nu^c-d_\nu^{ac}a_\mu^c\right)^2+2 F_{\mu\nu}^a f^{abc} a_\mu^b a_\nu^c\right]\,,
\ee
with $d_\mu^{ac}=\delta^{ac}\partial_\mu +f^{abc} \CA_\mu^b$ the background gauge-covariant derivative and where $F_{\mu\nu}^a$ is (\ref{fmunu}) evaluated for $A_\mu^a=\CA_\mu^a$. Because the path integral for QCD in this form contains flat directions, it is necessary to gauge-fix the theory. Following Weinberg's gauge fixing choice \cite{weinberg1995quantum}, this implies adding a gauge-fixing and ghost term of the form
\be
S_{\rm gf}=\frac{1}{2g_B^2}\int dx \left(d_\mu^{ac} a_\mu^c\right)^2\,,\quad
S_{\rm ghost}=\int dx \left[d_\mu^{ac} \bar c^{c} d_\mu^{ab}c^b+d_\mu^{ac} \bar c^c f^{adb} a_\mu^d c^b\right]\,,
\ee
to the action, where $\bar c,c$ are the Faddeev-Popov ghost fields.

The simplest solvable approximation of this theory is similar to the R0-level approximation in scalar field theory: just integrate out all quadratic terms in fields $a_\mu^a,\bar \psi,\psi,\bar c,c$ neglecting higher-order interactions. This corresponds to calculating the one-loop effective action, which similar to the case of scalar field theory corresponds to resumming an infinite number of perturbative contributions. Therefore, similar to scalar field theory, it corresponds to a non-perturbative calculation in QCD.

It should be pointed out that $\CA_\mu^a(x)$ is expected to take the same role as $\zeta_0$ in scalar field theory, e.g. corresponding to a saddle-point of the action. This immediately simplifies the calculation because $S_1[\CA]=0$ if $\CA_\mu^a$ is a saddle point. One finds
\be
\label{QCDpart}
Z=\int d\CA e^{-\frac{1}{2}{\rm Tr}\ln \left[\left(p^2 \delta^{ab}-2 i p_\alpha {\cal A}_\alpha^{ab}-\left({\cal A}^{2}\right)^{ab} \right) \delta_{\mu\nu}-2 {\cal F}_{\mu\nu}^{ab}\right]+{\rm Tr}\ln \left[p^2 \delta^{ab}-2 i p_\mu {\cal A}_\mu^{ab}-\left({\cal A}^{2}\right)^{ab}\right]+\frac{N_f}{2} {\rm Tr}\ln \slashed{D}^2}\,,
\ee
where ${\cal A}_\mu^{ab}=f^{acb}\CA_\mu^c$, ${\cal F}_{\mu\nu}^{ab}=f^{acb} F_{\mu\nu}^c$, ${\rm Tr}$ collectively denotes traces are over Lorentz, color, spinor and space--time indices where appropriate and some operators are represented in momentum space.

Evaluating $Z$ for arbitrary background field $\CA_\mu^a(x)$ is still too hard, so I am using a sub-set of possible functions $\CA_\mu^a(x)$, specifically those for which the background field-strength tensor $F_{\mu\nu}^a$ is constant and self-dual (equal Euclidean chromo-electric and chromo-magnetic fields), cf. the discussion in Ref.~\cite{Grable:2023eyj}. This amounts to setting
\be
\label{QCDconst}
\CA_\mu^a(x)=-\frac{1}{2}F_{\mu\nu}^a x_\nu\,.
\ee
Using then 
\be
\label{QCDtr}
  {\rm Tr}\ln {\cal O}=-\int_0^\infty \frac{ds}{s} K(s)\,, \quad K(s)={\rm Tr} e^{-s {\cal O}}\,,
  \ee
  the kernel function $K(s)$ can be written as a quantum-mechanical partition function
  \be
   K(s)={\rm tr}\int dx \left<x|e^{-s {\cal O}}|x\right>={\rm tr}\int {\cal D}x e^{-\CQ[x]}\,.
  \ee
  For the case of ghosts, gluons, and fermions one can find the ``actions'' $\CQ$ using results from \cite{Blau:1988iz,Strassler:1992zr,Reuter:1996zm} as in \cite{Grable:2023eyj}, specifically:
  \ba
   \label{sigmas}
    \CQ^{\rm ghost}&=&\int_0^{s} d\sigma \left[\frac{\dot x_\mu^2}{4}\delta^{ab}+ \dot{x}_\mu {\cal A}_\mu^{ab}\right]\,,\\
    \CQ^{\rm glue}&=&\int_0^{s} d\sigma \left[\frac{\dot x_\alpha^2}{4}\delta^{ab}\delta_{\mu\nu}+ \dot{x}_\alpha {\cal A}_\alpha^{ab} \delta_{\mu\nu}-2 {\cal F}_{\mu\nu}^{ab}\right]\,,\nonumber\\
    \CQ^{\rm fermion}&=&\int_0^{s} d\sigma \left[\frac{\dot x_\alpha^2}{4}+ i \dot{x}_\alpha \CA_\alpha^{a} t^a -\frac{i}{2}\sigma_{\mu\nu} F_{\mu\nu}^{a} t^a\right]\,.
  \ea
  In this form, the path integral representations for the kernel functions $K(s)$ look just like a finite-temperature quantum mechanics problem, and can be solved accordingly to give \cite{Grable:2023eyj}
  \ba
  K_{\rm ghost}(s)&=&\frac{\rm vol}{(4 \pi s)^{\frac{d}{2}}} \sum_{i=0}^3\frac{s^2\lambda_i}{2\sinh^2\left(\frac{s \sqrt{\lambda_i}}{2}\right)}\,,\nonumber\\
  K_{\rm glue}(s)&=& \frac{{\rm vol}\times d}{(4 \pi s)^{\frac{d}{2}}}\sum_{i=0}^3 \frac{s^2 \lambda_i \cosh\left(s \sqrt{\lambda_i}\right)}{2\sinh^2\left(\frac{s \sqrt{\lambda_i}}{2}\right)}\,,\nonumber\\
  K_{\rm fermion}(s)&=& \frac{{\rm vol}\times d}{(4 \pi s)^{\frac{d}{2}}}\sum_{i=1}^3 \frac{s^2 \lambda_{F,i} \cosh^2\left(\frac{s \sqrt{\lambda_{F,i}}}{2}\right)}{4\sinh^2\left(\frac{s \sqrt{\lambda_{F,i}}}{2}\right)}\,,
  \ea
  where $\lambda_i$ are the eigenvalues of the matrix ${\cal M}=-4 f^{acd} f^{deb} B^c B^e$ and $\lambda_{F,i}$ are the eigenvalues of the matrix ${\cal M}_F=4 t^a t^b B^a B^b$. Here $B^a$ is the (constant) chromomagnetic field corresponding to the entries of the self-dual constant field strength tensor $F_{\mu\nu}^a$.

  The partition function (\ref{QCDpart}) only involves a certain combination of the kernel functions
  \be
  K_{\rm sum}(s)=\frac{1}{2}K_{\rm glue}(s)-K_{\rm ghost}(s)-\frac{N_f}{2} K_{\rm fermion}(s)\,,
  \ee
  which for the specific choice of $N_f=4 N$ has been found to be free of IR divergencies \cite{Grable:2023eyj}. Any remaining divergences are UV divergencies, which can be extracted through dimensional regularization using $d=4-2\varepsilon$. Specifically, for the case $N_f=4 N$ one finds \cite{Grable:2023eyj}
  \be
   K_{\rm sum}=\frac{\rm vol}{(4\pi s)^{\frac{d}{2}}}\left[(d-2)
\sum_{i=1}^3
\left(\frac{(d-2)s^2 \lambda_i}{4\sinh^2\left(\frac{s \sqrt{\lambda_i}}{2}\right)}
  -\frac{N_f d}{2}\frac{s^2\lambda_{F,i}}{4 \sinh^2\left(\frac{s \sqrt{\lambda_{F,i}}}{2}\right)}\right)\right]\,,
   \ee
   so that the traces of the operators in (\ref{QCDtr}) can be evaluated in terms of $\zeta$-functions \cite[25.5.9]{NIST:DLMF}. After the dust settles, one has
   \be
   Z=\int d\CA e^{-{\rm vol}\times {\cal F}}\,,
   \ee
   with the one-loop QCD effective potential
   \be
   {\cal F}=\frac{B^a B^a}{g_B^2}+\frac{-N B^a B^a}{(4\pi^2)\varepsilon}+\frac{12 N B^a B^a\zeta^\prime(-1)}{(4\pi)^2}
      +\sum_{i=1}^3\left(\frac{\lambda_i\ln \frac{\bar \mu^2e^{-1}}{\sqrt{\lambda_i}}-N_f \lambda_{F,i}\ln \frac{\bar\mu^2 e^{-\frac{1}{2}}}{\sqrt{\lambda_{F,i}}}}{96\pi^2}\right)\,,
      \ee
      where the relation between $\CA$ and $B$ is provided by (\ref{QCDconst}) and the eigenvalues $\lambda_i,\lambda_{F,i}$ are functions of $B$. At this point, I again wanted to stress the similarity to the scalar field theory calculation at the R0 level, notably (\ref{ZR0}) and (\ref{FR0}). In the large volume limit, the QCD partition function is given exactly as the saddle point of the action
      \be
      \ln Z = -{\rm vol}\times F[\bar B^a]\,,
      \ee
      with $\bar B^a$ the location of the saddle (there is a Jacobian when changing the integration from $\CA^a_\mu$ to $B^a_\mu$, but the contribution from this Jacobian is suppressed in the large volume limit). Similar to the case of scalar field theory, the 0-point function for QCD is finite if the coefficient of $B^a B^a$ in ${\cal F}$ is finite. This gives the non-perturbative renormalization condition for QCD with $N_f=4N$ fermion flavors as
      \be
      \label{R0QCD}
      \frac{1}{g_{B}^2}-\frac{N}{(4\pi)^2 \varepsilon}=\frac{1}{g_R^2(\bar\mu)}\,,
      \ee
      which implies a running QCD coupling of the form
      \be
      g^2_R(\bar\mu)=\frac{(4\pi)^2}{N \ln \frac{\bar \mu^2}{\Lambda_{\msbar}^2}}\,.
      \ee
      Note that this running coupling matches the one-loop perturbative running for QCD with $\beta_0=\frac{11 N}{3}-\frac{2 N_f}{3}$ and $N_f=4 N$. The bare Yang-Mills coupling fulfills
      \be
      \label{nonpert}
      \lim_{\varepsilon\rightarrow 0}g_B^2=\lim_{\varepsilon\rightarrow 0}\frac{1}{\frac{1}{g_R^2(\bar\mu)}+\frac{N}{(4\pi)^2\varepsilon}}\rightarrow 0\,,
      \ee
      whereas \textit{first} expanding $g_B^2$ for small coupling $g_R^2$ and \textit{then} taking the regulator to zero gives the perturbative result
      \be
      \label{QCDpert}
      \lim_{\varepsilon\rightarrow 0}g_{B,{\rm pert}}^2=\lim_{\varepsilon\rightarrow 0}\left[g_R^2(\bar\mu)-\frac{N g_R^4}{(4\pi)^2\varepsilon}+{\cal O}(g_R^6)\right]\rightarrow -\infty\,,
      \ee
      which differs qualitatively from the non-perturbative alternative renormalization (\ref{nonpert}).

      After non-perturbative renormalization (\ref{R0QCD}), one obtains
      \be
      {\cal F}=\frac{N B^a B^a}{(4\pi)^2}\ln \frac{\bar\mu^2}{\Lambda_{\msbar}^2}+\frac{12 N B^a B^a\zeta^\prime(-1)}{(4\pi)^2}
      +\sum_{i=1}^3\left(\frac{\lambda_i\ln \frac{\bar \mu^2e^{-1}}{\sqrt{\lambda_i}}-N_f \lambda_{F,i}\ln \frac{\bar\mu^2 e^{-\frac{1}{2}}}{\sqrt{\lambda_{F,i}}}}{96\pi^2}\right)\,,
      \ee
      and it is straightforward to check that ${\cal F}$ is independent from the fictitious renormalization scale $\bar\mu$. In full analogy to the scalar field theory case (\ref{R0saddles1}), ${\cal F}$ has a trivial saddle $\bar B^a=0$ as well as non-trivial saddles $\bar B^a\neq 0$. The lower free energy is found for non-trivial saddles of full rank, e.g. for
      \be
      \bar B^a=\delta^{a8}e^{-12 \zeta^\prime(-1)-\frac{5}{6}}{2^{-\frac{8}{9}}3^{\frac{5}{6}}}\Lambda^2_{\msbar}\simeq 4.27 \Lambda^2_{\overline{\rm MS}}\delta^{a8}\,.
      \ee
      The explicit (finite!) result for the 0-point function for this non-trivial saddle is
      \be
      {\cal F}[\bar B^a]=-\frac{3 \bar B^a \bar B^a}{32\pi^2}\,.
      \ee

      For the 2-point functions of the gauge fields, fermions and ghosts, one notices that similar to the case of scalar field theory (\ref{2pointR0}) these are UV finite as long as the saddle location $\bar B^a$ is finite. Therefore, the non-perturbative renormalization condition (\ref{R0QCD}) automatically renders all 2-point functions of the theory UV-finite. However, it should be stressed that while the 2-point functions are UV-finite, they still do contain IR divergencies at this level of resummation \cite{Savvidy:2022jcr,Grable:2023eyj}. If they were automatically IR finite, the mass gap problem for Yang Mills would be solved, and I would enjoy the luxury of being able to refuse to sit on boring university committees. However, renormalization of quantum field theory is concerned with UV divergencies, and these are taken care of by (\ref{R0QCD}).

      Similarly to the case of scalar field theory, higher-order n-point functions require going beyond a simple R0-level resummation scheme, which so far is not available for QCD. However, it should be pointed out that the non-perturbative renormalization condition (\ref{R0QCD}) renders both the 0-point and 2-point functions of $N_f=4N$ QCD finite, whereas in perturbation theory this requires both coupling and wave-function counterterms.

      An interesting observable for $N_f=12$ QCD to evaluate is the expectation value of the trace of the Polyakov loop in the presence of the non-trivial background saddle $\bar B^a$. One can explicitly evaluate this observable using (\ref{QCDconst}) and finds
      \be
      {\rm tr} \int d^3 \vec{x} {\cal P} e^{i\oint d\tau t^a \CA_0^a(\tau,\vec{x})} = 0\,,
        \ee
        where ${\cal P}$ denotes path ordering. Vanishing Polyakov loop expectation value is typically associated with confinement in QCD, so the above result suggests that QCD with $N_f=12$ massless flavors is confining. This is consistent with results from lattice QCD simulations from Ref.~\cite{Fodor:2017gtj}, though perturbative and other more recent lattice studies found that the theory is instead infrared conformal\cite{DeGrand:2015zxa,DiPietro:2020jne,Peterson:2024pyt}.

        I leave the resolution of the question of conformality in $N_f=12$ QCD for future studies, only pointing out here that non-perturbative renormalization of the theory allows for predictive non-perturbative calculations.
            
      \section{Summary and Conclusions}

      In this work, I have considered alternative renormalization procedures for scalar field theory and QCD in dimensional regularization. I provided explicit results that showed that this alternative renormalization procedure leads to finite n-point correlation functions even though it differs qualitatively from the usual perturbative renormalization approach.

      In particular, there is no wave-function counterterm in the non-perturbative approach.

      Furthermore, the bare coupling constant in non-perturbative renormalization differs qualitatively from the perturbative scheme in that it tends to zero in the limit of vanishing regulator.

      In the case of QCD, the non-perturbative renormalization seems much more similar to lattice QCD \cite{Karsch:2001cy} than to perturbative calculations \cite{vanRitbergen:1997va}. Specifically, the expression for the bare coupling constant $g_B$ in non-perturbative renormalization is qualitatively consistent with the dependence of the bare lattice coupling on the regulator, see e.g. Ref.~\cite{Boyd:1996bx}.

      In the case of scalar field theory, the results further strengthen the case for revisiting the claims of quantum triviality in four dimensions \cite{Aizenman:2019yuo,Romatschke:2023sce}. However, further advances on lattice field theory along the lines of \cite{Lawrence:2022afv,Romatschke:2023fax} or experimental advances along the lines of \cite{Rotter:2016,Xu:2016} would be needed to unambiguously settle this question.

      To conclude, perturbative renormalization is not the only option for dealing with UV divergencies in quantum field theory. Non-perturbative alternative schemes exist which seem to work just as well as their perturbative counterparts, yet with fewer free parameters.

\section*{Acknowledgments}

This work was supported by the Department of Energy, DOE award No DE-SC0017905. I am grateful to Joaquin Drut, Hendrik van Hees, Thimo Preis and Zsolt Szep for helpful comments and questions, and to Willie Su for pointing out typos in the appendix of this manuscript.

\begin{appendix}
  \section{Semi-analytic evaluation of divergent integrals in dimensional regularization}

  Some integrals encountered in the main text are not routinely encountered in perturbative quantum field theory, so their evaluation in dimensional regularization may not be standard. For this reason, I provide a quick guide on how to evaluate such integrals semi-analytically in this appendix.

  To get started, let me consider a simple divergent integral $I_1$ in dimensional regularization, and pretend that I do not know that it corresponds to a representation of the $\Gamma$-function:
  \be
  \label{actual1}
  I_1=\mu^{2\varepsilon}\int \frac{d^d k}{(2\pi)^d}\frac{1}{(k^2+m^2)^2}=\left(\frac{4\pi \mu^2}{m^2}\right)^{\varepsilon}\frac{\Gamma\left(\varepsilon\right)}{16\pi^2}\,.
  \ee
  The integrand does not depend on the angular variables, so that using the volume of a hypersphere in $d=4-2\varepsilon$ dimensions one finds
  \be
  I_1=\frac{(4\pi\mu^2)^{\varepsilon}}{8\pi^{2}\Gamma\left(2-\varepsilon\right)}\int_0^\infty dk\, \frac{k^{3-2\varepsilon}}{(k^2+m^2)^2}\,.
  \ee
  Introducing the new variable $x=k^2$, and scaling $x\rightarrow x m^2$, this becomes
   \be
  I_1=\left(\frac{4\pi \mu^{2}}{m^2}\right)^{\varepsilon}\frac{1}{16\pi^{2}\Gamma\left(2-\varepsilon\right)}\int_0^\infty dx\, \frac{x^{1-\varepsilon}}{(x+1)^2}\,.
  \ee
  Pretending not to know the exact form of the integral, one nevertheless knows that it is logarithmically divergent, so I split it into two parts:
  \be
  \int_0^\infty dx\, \frac{x^{1-\varepsilon}}{(x+1)^2}=\int_0^R dx\, \frac{x^{1-\varepsilon}}{(x+1)^2}+\int_R^\infty dx\, \frac{x^{1-\varepsilon}}{(x+1)^2}\,,
  \ee
  with regulator $R$, so that the integration $x\in [0,R]$ is always convergent for any finite $R$. For the integration $x>R$, one can extract the divergent contribution by performing an Taylor expansion in $\frac{1}{x}$, so that
  \ba
  \int_R^\infty dx\, \frac{x^{1-\varepsilon}}{(x+1)^2}&=&\int_R^\infty dx\, \left(\frac{x^{1-\varepsilon}}{(x+1)^2}-x^{-1-\varepsilon}\right)+\int_R^\infty dx\, x^{-1-\varepsilon}\,,\nonumber\\
  &=&\int_R^\infty dx\, \left(\frac{x^{1-\varepsilon}}{(x+1)^2}-x^{-1-\varepsilon}\right)-\left.\frac{x^{-\varepsilon}}{\varepsilon}\right|_{x=R}^\infty\,.
  \ea
  In dimensional regularization, the integral is evaluated \textit{first} at $x=\infty$ for sufficiently high $\epsilon>0$ such that the result is finite. \textit{Then} the limit $\epsilon\rightarrow 0$ is performed. The remaining contributions for $I_1$ are all finite for $\epsilon\rightarrow 0$, so that they can be evaluated numerically. One finds
  \be
  I_1=\lim_{\varepsilon\rightarrow 0}\left(\frac{4\pi\mu^{2}}{m^2}\right)^{\varepsilon}\frac{1}{16\pi^{2}\Gamma\left(2-\varepsilon\right)}\left(\int_0^R dx\, \frac{x^{1}}{(x+1)^2}+\int_R^\infty dx\, \left(\frac{x^{1}}{(x+1)^2}-x^{-1}\right)+\frac{R^{-\varepsilon}}{\varepsilon}\right)\,.
  \ee
  The remaining integrals are evaluated numerically for different choices of $R$, verifying that the result is independent of $R$. Specifically, one finds e.g. for various values of $R$ to within floating point precision
  \ba
   R=3: &\quad& \int_0^R dx\, \frac{x^{1}}{(x+1)^2}+\int_R^\infty dx\, \left(\frac{x^{1}}{(x+1)^2}-x^{-1}\right)-\ln R=-1.\,,\nonumber\\
   R=4: &\quad& \int_0^R dx\, \frac{x^{1}}{(x+1)^2}+\int_R^\infty dx\, \left(\frac{x^{1}}{(x+1)^2}-x^{-1}\right)-\ln R=-1.\,,\nonumber\\
    R=5: &\quad& \int_0^R dx\, \frac{x^{1}}{(x+1)^2}+\int_R^\infty dx\, \left(\frac{x^{1}}{(x+1)^2}-x^{-1}\right)-\ln R=-1.\,.\nonumber\\
  \ea

  Therefore, the semi-analytical result for $I_1$ in dimensional regularization becomes
  \be
  I_1=\lim_{\varepsilon\rightarrow 0}\left(\frac{4\pi\mu^{2}}{m^2}\right)^{\varepsilon}\frac{1}{16\pi^{2}\Gamma\left(2-\varepsilon\right)}\left(\frac{1}{\varepsilon}-1.\right)\simeq \frac{1}{16\pi^2}\left(\frac{1}{\varepsilon}+\ln \frac{\bar\mu^2}{m^2}\right)\,,
  \ee
  where $\bar\mu^2=4\pi\mu^2 e^{-\gamma_E}$, matching the expected result (\ref{actual1})

  The same technique can be applied to more complicated integrals, e.g.
  \be
  I_2=\mu^{2\varepsilon}\int \frac{d^dk}{(2\pi)^d}\ln \left[c+2 \sqrt{\frac{k^2+4 m^2}{k^2}}{\rm atanh}\sqrt{\frac{k^2}{k^2+4 m^2}}\right]\,,
  \ee
  with $c>-2$ an arbitrary constant. Using the same steps as for $I_1$, one finds
  \be
  I_2=\left(\frac{4 \pi \mu^2}{m^2}\right)^{\varepsilon}\frac{m^4}{16\pi^2\Gamma\left(2-\varepsilon\right)}\int_0^\infty dx x^{1-\varepsilon}\ln \left[c+2 \sqrt{\frac{x+4}{x}}{\rm atanh}\sqrt{\frac{x}{x+4}}\right]\,,
  \ee
  where the remaining integral is again split using an arbitrary regulator $R$.
  One finds
  \be
  I_2=\left(\frac{4 \pi \mu^2}{m^2}\right)^{\varepsilon}\frac{m^4}{16\pi^2\Gamma\left(2-\varepsilon\right)}\left(I_{2L}+I_{2H}+I_{D1}+I_{D2}+I_{D3}\right)\,,
  \ee
  where
  \ba
  I_{2L}&=&\int_0^R dx x^{1}\ln \left[c+2 \sqrt{\frac{x+4}{x}}{\rm atanh}\sqrt{\frac{x}{x+4}}\right]\,,\\
  I_{2H}&=&\int_R^\infty dx x^{1}\left\{\ln \left[\frac{c+2 \sqrt{\frac{x+4}{x}}{\rm atanh}\sqrt{\frac{x}{x+4}}}{c+\ln x}\right]-\frac{2}{x}\left(1+\frac{1-c}{c+\ln x}\right)\right.\nonumber\\
  &&\left.-\frac{c-2-(3+2c)\ln x-4 \ln^2 x}{x^2\left(c+\ln x\right)^2}\right\}\,,\\
  I_{D1}&=&\int_R^\infty dx x^{1-\varepsilon}\ln\left(c+\ln x\right)\,,\\
  I_{D2}&=&2 \int_R^\infty dx x^{-\varepsilon}\left(1+\frac{1-c}{c+\ln x}\right)\,,\\
  I_{D3}&=&\int_R^\infty dx x^{-1-\varepsilon}\frac{c-2-(3+2c)\ln x-4 \ln^2 x}{\left(c+\ln x\right)^2}\,.
  \ea
  The divergent pieces $I_{D1},I_{D2},I_{D3}$ are again calculated analytically, and then evaluated for $x\rightarrow \infty$ using sufficiently large $\varepsilon>0$ to render them convergent. Taking $\varepsilon\rightarrow 0$ as a second step, they explicitly become
  \ba
  I_{D1}&=&\frac{1}{2}\left[-R^2\ln\ln \left(R e^{c}\right)+e^{-2 c}{\rm Ei}\left(\ln R^2e^{2c}\right)\right]\,,\\
  I_{D2}&=&-2\left[R+(1-c)e^{-c}{\rm Ei}\left(\ln R e^{c}\right)\right]\,,\\
  I_{D3}&=&-\frac{4}{\varepsilon}+4\ln R+(3-6c)\left(\gamma_E+\ln\varepsilon+\ln\ln (R e^{c})\right)-\frac{2 (c-1)^2}{\ln R e^{c}}\,.
  \ea
  with ${\rm Ei}$ denoting the exponential integral function.
  
  Evaluating the remaining integrals numerically for given $c,R$, one finds for instance for $c=-1$:
  \be
  I_2=\left(\frac{4 \pi \mu^2}{m^2}\right)^{\varepsilon}\frac{m^4}{16\pi^2}\left(-\frac{4}{\varepsilon}-4 +13 \gamma_E+9\ln \varepsilon-0.278387\ldots \right)
  \ee
  
  \end{appendix}

\bibliography{PT}
\end{document}